\documentclass[traditabstract,longauth,letter]{aa}
\usepackage{txfonts}
\usepackage{natbib}
\usepackage{graphicx}
\bibpunct{(}{)}{;}{a}{}{,}

\newcommand{\CSO}{\textrm{\CSO}}

\newcommand{\HH}    {\mbox{H$_2$}}           

\newcommand{\HCOp}  {\mbox{HCO$^{+}$}}       

\newcommand{\Jone}{\mbox{$J$=1--0}}

\newcommand{\emm}[1]{\ensuremath{#1}}   
\newcommand{\emr}[1]{\emm{\mathrm{#1}}} 
\newcommand{\unit}[1]{\emm{\, \emr{#1}}}

\newcommand{\pscm}{\unit{cm^{-2}}}

\newcommand{\kms}   {\unit{km\,s^{-1}}}

\newcommand{\Msun}  {\unit{M_\odot}}

\renewcommand{\deg}{\emm{^\circ}}

\def\cc{\ifmmode{\,{\rm cm}^{-3}}\else{$\,{\rm cm}^{-3}$}\fi}
\def\cq{\ifmmode{\,{\rm cm}^{-2}}\else{$\,{\rm cm}^{-2}$}\fi}
\def\eccs{\ifmmode{\,{\rm erg}\,{\rm cm}^{-3} {\rm s}^{-1}}\else{$\,{\rm
erg}\,{\rm cm}^{-3} {\rm s}^{-1}$}\fi}
\def \Cp{\ifmmode{\rm C^+}\else{$\rm C^+$}\fi} 
\def \CHtwp{\ifmmode{\rm CH_2^+}\else{$\rm CH_2^+$}\fi}
\def \CHthp{\ifmmode{\rm CH_3^+}\else{$\rm CH_3^+$}\fi}
\def \CHp{\ifmmode{\rm CH^+}\else{$\rm CH^+$}\fi}
\def \thCHp{\ifmmode{\rm ^{13}CH^+}\else{$\rm ^{13}CH^+$}\fi}
\def \twCHp{\ifmmode{\rm ^{12}CH^+}\else{$\rm ^{12}CH^+$}\fi}
\def\wat{\ifmmode{\rm H_2O}\else{$\rm H_2O$}\fi} 

\begin{document}
%
%
  \title{Strong CH$^+$ \Jone\ emission and absorption in DR21
  \thanks{\emph{Herschel} is an ESA space observatory with science instruments provided by 
European-led Principal Investigator consortia and with important participation from NASA.}}

   \authorrunning{E. Falgarone, V. Ossenkopf, M. Gerin et al.}
   \titlerunning{Strong CH$^+$ \Jone\ emission and absorption in DR21}

\author{E. Falgarone\inst{1},
V. Ossenkopf\inst{2,3}, 
M. Gerin\inst{1},
P. Lesaffre\inst{1},
B. Godard\inst{4}, 
J. Pearson\inst{5},
S.~Cabrit\inst{6},
Ch.~Joblin\inst{7},
A.~O.~Benz\inst{8},
F.~Boulanger\inst{4},
A.~Fuente\inst{9},
R.~G\"usten\inst{10},
A. Harris\inst{11},
T.~Klein\inst{10},
C.~Kramer\inst{12},
S.~Lord\inst{13},  
P.~Martin\inst{14},
J.~Martin-Pintado\inst{15},
D.~Neufeld\inst{16},
T.~G.~Phillips\inst{17},
M. R\"ollig\inst{2},
R. Simon\inst{2}, 
J. Stutzki\inst{2}, 
F.~van der Tak\inst{3,18},
D.~Teyssier\inst{19},
H.~Yorke\inst{5},
N.~Erickson\inst{20}, 
M.~Fich\inst{21},
W.~Jellema\inst{3},
A.~Marston\inst{19},
C.~Risacher\inst{3},
M.~Salez\inst{6},
F. ~Schm\"ulling\inst{2}
}
\institute{LERMA, CNRS, Observatoire de Paris \& Ecole Normale Sup\'erieure, 24 rue Lhomond, 75005 Paris, France 
\and
I. Physikalisches Institut der Universit\"at 
zu K\"oln, Z\"ulpicher Stra\ss{}e 77, 50937 K\"oln, Germany
\and
SRON Netherlands Institute for Space Research, P.O. Box 800, 9700 AV 
Groningen, Netherlands
\and
Institut d'Astrophysique Spatiale, CNRS and Universit\'e Paris-Sud, B\^at. 121, 91405 Orsay Cedex, France
\and
Jet Propulsion Laboratory, Caltech, 4800 Oak Grove Drive, Pasadena CA 91109-8099, USA
\and 
LERMA, CNRS \& Observatoire de Paris , 61 av. de l'Observatoire, Paris, France  
\and
CESR, CNRS and Universit\'e Paul Sabatier, 9 avenue du Colonel Roche, 31062 Toulouse, France
\and
Institute for Astronomy, ETH Z\"urich, 8093 Z\"urich, Switzerland
\and
Observatorio Astron\'omico Nacional (OAN), Apdo. 112, 28803 Alcal\'a de Henares (Madrid), Spain
\and
Max-Planck-Institut f\"ur Radioastronomie, Auf dem H\"ugel 69, 53121, Bonn, Germany
\and 
Astronomy Department, University of Maryland, College Park, MD 20742, USA
\and
Instituto de Radio Astronom\'ia Milim\'etrica (IRAM), Avenida Divina Pastora 7, Local 20, 18012 Granada, Spain
\and
IPAC/Caltech, MS 100-22, Pasadena, CA 91125, USA
\and 
Department of Astronomy and Astrophysics, University of Toronto, 60 St. George Street, Toronto, ON M5S 3H8, Canada
\and
Centro de Astrobiolog\'ia, CSIC-INTA, 28850, Madrid, Spain
\and
Department of Physics and Astronomy, John Hopkins University, 3400 North Charles Street, Baltimore, MD 21218, USA
\and
California Institute of Technology, 320-47, Pasadena, CA  91125-4700, USA
\and
Kapteyn Astronomical Institute, University of Groningen, PO box 800, 9700 AV Groningen, Netherlands
\and
European Space Astronomy Centre, Urb. Villafranca del Castillo, P.O. Box 50727, Madrid 28080, Spain
\and
University of Massachusetts, Astronomy Dept.,  Amherst, MA 01003-9305  U.S.A.
\and 
Department of Physics and Astronomy, University of Waterloo, Canada N2L 3G1
}


\date{Received / Accepted }

%
%
\abstract
{We report the first detection of the ground-state rotational transition of the
methylidyne cation CH$^+$ towards the massive star-forming region DR~21 with
the HIFI instrument onboard the \emph{Herschel} satellite. The line
profile exhibits a broad emission line, in addition to
two deep and broad absorption features associated with the DR~21 molecular ridge and 
foreground gas.
These observations allow us to determine a \twCHp\Jone\ line frequency of
$\nu$=835137$\pm3$~MHz, in good agreement with 
 a recent experimental determination. We estimate the CH$^+$ column density  to be 
a few $10^{13}$ cm$^{-2}$ in the gas seen in emission, and $> 10^{14}$ cm$^{-2}$ 
in the components responsible for the absorption, which is indicative of a high 
line of sight average abundance [\CHp]/[H]$> 1.2\times 10^{-8}$.
We show that the \CHp\ column densities agree well with the predictions of 
state-of-the-art C-shock models in dense UV-illuminated gas for the emission line, 
and with those of turbulent dissipation models in diffuse gas for the absorption 
lines.}
%
%


\maketitle


%
%
\section{Introduction}
%
%
The methylidyne ion CH$^+$ was among  the first molecules to be
detected in the interstellar medium (ISM) \cite{douglas41}. This reactive ion
is prevalent in the diffuse ISM with column densities 
several orders of magnitude 
above the predictions of UV-driven equilibrium models  
\citep[see references in][]{godard09}.  
Apart from the strong emission lines of CH$^+$ 
 detected in both the envelope of the Red Rectangle \cite{hobbs04} and  NGC~7027 \cite{cernicharo}, 
all other observations are absorption lines detected at visible wavelengths 
in the spectra of nearby stars.
%
\CHp\ is a light molecule and its rotational lines 
lie at submillimeter and far infrared wavelengths. The exact frequencies
of the rotational transitions of CH$^+$ have remained elusive for a long time because
of the extreme reactivity of CH$^+$ and the difficulty in
isolating it in the laboratory \cite{pearson06}. 
Recent laboratory measurements have led to
 $\nu=835.137498(20)$ GHz for the ground-state transition
(Amano, 2010).
Ground-based astronomical detection of \twCHp(1-0) is prevented
by its proximity to a strong atmospheric line of water vapor.
The ground-state frequency
 of the isotopologue \thCHp, redshifted by $\sim 5$ GHz, has
superior sky transmission. It has been detected at the
Caltech Submillimeter Observatory towards several massive star-forming regions
of the inner Galaxy \citep[Falgarone et al. in prep., ][]{falgarone05}. The \twCHp\ 
abundances averaged along these long lines-of-sight ({\it los}), 
confirm the high abundances of this species 
inferred from visible observations in the local ISM, [\CHp]/[H]$\sim$ 8$\times 10^{-9}$
on average.
In this paper, we report the detection of the 
\twCHp\Jone\ transition towards the massive star-forming region DR21, presented in Sect. 2.
The HIFI observations are described in Sect. 3. The results, given in Sect. 4, are 
compared to models in Sect. 5. 

\section{The DR~21 region}
The massive star-forming region DR~21 is located in the Cygnus X 
complex at an average distance of 1.7 kpc, that of the 
Cyg OB2 stellar association  \citep[see reviews by][]{jakob,schneider}. 
The DR~21 molecular core is one of the most massive cores in the Galaxy 
(2.5$\times 10^4$ \Msun\ at the 1~pc-scale) 
\citep{kirby}. It is located in front of the main DR~21 HII region comprising 
five compact HII regions 
 and more diffuse ionised gas \citep{prr,wilson,cyganowski}.

The star-forming region DR~21 itself is known to host one of the most powerful molecular outflows
shining in vibrationally excited H$_2$ \citep[e.g.,][]{garden}. The outflow
source is heavily extinguished by more than 100 magnitudes, and
its location is not accurately  known \citep{wilson}.
High velocity HI (up to 90 kms$^{-1}$) 
associated with the molecular outflow has been detected
by the VLA \citep{russell}, indicating that the molecular outflow
may be driven by this atomic jet. 
The outflow shows up as broad wings in CO and HCO$^+$ lines \cite{kirby}.
At the position of the HIFI beam, the $v = 1-0$ S(1) 
\HH\ emission line has a similar broad width  \cite{nadeau,cruz07} to that of
the SiO emission \citep{motte07}.

Some line profiles imply that there is foreground gas, 
which is probably associated with the Cygnus X complex
\cite{jakob,schneider}. A foreground component associated with W75N shows up as
weak emission lines at $v_{LSR} \sim 8$ km s$^{-1}$ 
in low excitation transitions of CO and HCO$^+$ or atomic carbon, and
in the atomic oxygen fine-structure line, with 
$N({\rm O}) > 5 \times 10 ^{18}$ cm$^{-2}$ \cite{poglitsch}. The total hydrogen 
column density of this  foreground material is estimated to be
$N({\rm H}) \sim 1.4 \times 10^{22}$ cm$^{-2}$ for an elemental abundance ratio 
[O]/[H] =$3.45 \times 10^{-4}$ \cite{oliveira}, in coherence with
$N({\rm H}) = (1.3 \pm 0.4) \times 10^{22}$ cm$^{-2}$ inferred from K extinction 
to DR21 \citep{marshall06}.
This component is also detected in atomic hydrogen, which indicates that there is 
saturated absorption of between -5 and 18 \kms\ \cite{roberts}. Its
estimated HI column density, if $T_s = 20K$, is $N$(HI) $> 1.5 \times 10^{21}$
cm$^{-2}$.  


%
%
%
\section{HIFI observations}

\begin{figure}
\centering
\rotatebox{270}{
\includegraphics[width=0.33\textwidth,angle=90]{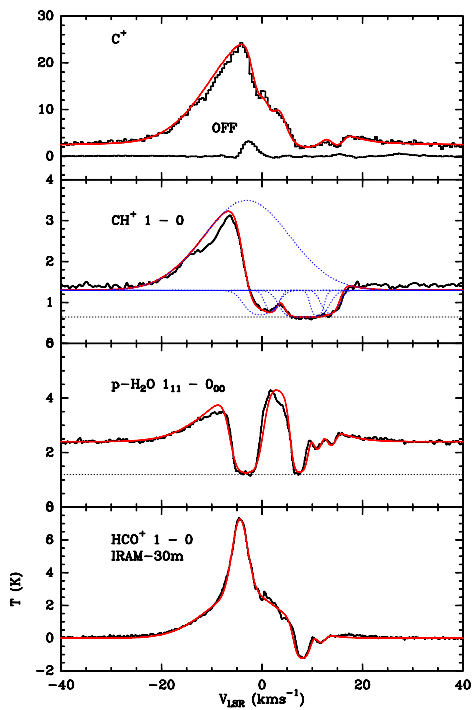}
}
\caption{[CII], \CHp(1-0) and p-\wat\ \emph{Herschel}/HIFI spectra (top panels) and
\HCOp(1-0) IRAM-30m spectrum  (bottom panel) \citep{schneider}. The HIFI spectra are taken in
DSB, hence saturated lines reach to half the continuum level (dotted lines).
The weak line in the C$^+$ panel is that of an intermediate
OFF position. The CH$^+$ (J=1-0) spectrum is shown
assuming a rest frequency of 835137~MHz. The Gaussian velocity components involved in the  
empirical model of the \CHp\ profile are shown (dotted blue curves) as well as the 
resulting models for all lines (red curves) (see Sect. 4).}
\label{fig:data}
\end{figure}

All FIR spectra presented here were obtained in the course of performance
verification observations with the HIFI instrument \citep{pilbratt10,degraauw10}. 
Since their main goal
was to demonstrate the functionality and performance of the different
observing modes, the spectra were taken with a large variety of
observing strategies.
Most observations were only single-point observations towards the
central position of the DR21 H{\sc II} region at RA=20h39m01.1s,
Dec=42$\deg$19$'$43.0$''$ (J2000). At the frequencies around 1~THz
discussed here, the \emph{Herschel} beam covers about 20$''$ HPBW 
(or 0.16~pc at 1.7~kpc). 
Here, we only present data from the wideband spectrometer (WBS)   
that provides a resolution
of 1.1~MHz, corresponding to 0.2~km/s (at 1900~GHz) and 0.4~km/s (at 835~GHz). 

The CH$^+$ spectrum at 835~GHz was obtained in a single-point load-chop
observing mode without OFF reference. It uses the
internal cold load as reference, having
the advantage that we can exclude any
self-chopping effects in the line profiles.
The mode
has, however, the disadvantage that the different optical paths towards
the sky and the internal calibration load sometimes lead to standing wave
differences that are detected as baseline ripples. We subtracted these manually using 
the HifiFitFringe pipeline tool. 
The total integration time on the source was 48~s leading 
to a noise level of 0.1~K for the combination of the spectra
from both polarisations at the WBS resolution.

The H$_2$O $1_{11}-0_{00}$ (1113~GHz)
line was observed 
in the frame of
a spectral scan in a load-chop mode and an on-source 
integration time of 30~s. The baseline was calibrated on  
 a distant OFF position (20h37m10s, 42$\deg$37$'$00$''$) because the
first selected OFF position, closer to the ON, retained some signal (see Fig. 1).  
The [C{\sc II}] line was detected in one of the first double-beam switch
raster mapping observations in which neither the logic of the mapping mode
worked as expected nor the pointing of the instrument was
yet known. The instrument pointing was measured afterwards
and the map coordinates were corrected by hand to include the correct
offsets. A comparison of the integrated [C{\sc II}] 
intensities in the shifted map 
with the MSX 8~$\mu$m band shows a very good agreement, confirming 
the accuracy of the correction.
The observations were taken on a
7$''$ raster that included a 14~s on-source integration time at each point.
To obtain a spatial resolution comparable to that of the other two lines, the 
map was convolved with a Gaussian beam of 20$''$ at the DR~21 central position. 




%
%
\section{Results of the line profile analysis}
%
%

The HIFI spectra are displayed in Fig. \ref{fig:data} with, for comparison, an IRAM-30m 
spectrum of \HCOp(1-0) \citep{schneider}. 
The line profiles exhibit both complex
emission and absorption  because of the multiplicity of velocity components and gas 
physical conditions within the beam.
The line integrated areas in emission and absorption 
are remarkably comparable. This would lead to a weak or non-detection 
with low spectral resolution instruments.
The dashed  lines in the \CHp\ and p-\wat\ spectra, at about
 half the continuum level,
reveal the broad velocity ranges across which the absorption lines are saturated.

\begin{figure}
  \centering
\includegraphics[width=0.4\textwidth,angle=0]{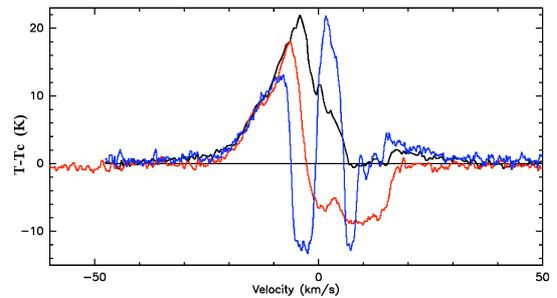}
\caption{Superimposition of the CH$^+$ (red), \Cp\ (black), and p-\wat\ (blue) spectra (from
which the continuum has been subtracted) to illustrate the good coincidence of the 
blue wing of 
the line  profile when the \CHp(1-0) frequency is $\nu=835.137$~GHz. We note also
the coincidence of the absorption features in \CHp(1-0) and \Cp. }
\label{fig:3spec}
\end{figure}

The \CHp(1-0), \Cp\ and p-\wat\ spectra, plotted after their continuum has been subtracted, 
are superimposed in Fig. \ref{fig:3spec} for the frequency of the \CHp\Jone\ line 
set to 835.137 GHz. We propose in Sect. 5 that the broad line emission originates from a
shock associated with the outflow, and we show that the three species, 
\CHp, \Cp\ , and \wat\ are connected with each other in the shock chemistry.
The good  agreement (within $\pm$ 1 \kms) between the blue 
wing of the emission line profiles of these three species allows us to adopt 
$\nu=835137 \pm 3$~MHz as the rest frequency of the \twCHp(1-0) line, which  
is remarkably consistent with the experimental value measured by 
Amano (2010).


\begin{table}
\caption{Empirical models of the broad emission profiles}
\label{tab:line}
\begin{tabular}{lllllll}
\hline\hline
Line & $v$ & $\Delta v$ & $T$ & & & \\
           & \kms\ &   \kms\ & K & & & \\
\hline
 CH$^+$  & -3.0 & 20.0 & 2.2 & & & \\ 
       p-H$_2$O & -1.0 & 20.0 & 2.16 & & & \\
       C$^+$ &   -1.8   & 20.0  & 22 & & &  \\
              &  0    & 80.0  & 1.0 & & & \\
       HCO$^+$ & -3.6 & 17.0 & 2.8 & & &   \\
\hline
\end{tabular}
\end{table}

The CH$^+$ \Jone\ line profile can be decomposed into a broad emission
line and a series of absorption features. We used information
from the line profiles of CO and HCO$^+$ \Jone\ (see Fig. \ref{fig:data}),
the spectra of atomic carbon \cite{jakob}, and other HIFI spectra
\citep{ossenkopf1,vdt} to constrain the velocity and linewidth of
these components. 
Since the \thCHp\ line has not been observed, the actual \CHp\ emission profile
is unknown. We had to rely on
  the symmetric shapes of the CO emission lines to model the broad emission wings
with  Gaussians adjusted on the blue wings of the lines (Table 1).  
  Over the velocity range -5 to 17 \kms, not only the continuum level but also the 
core line emission is absorbed by intervening gas. 
The broad emission lines ($\Delta v=20$~\kms, see Table 1) correspond to
  the emission associated with the outflow shock 
(see Sect. 5). A still broader component
in the [CII] line has a width similar to that of the 
H66$\alpha$  recombination lines \cite{prr}.
Self-absorption features in the range $\sim -3$ to 2 \kms, 
caused by the close environment of the DR21 core, are visible in CO, HCO$^+$(4-3), 
and other dense gas tracers (e.g., NH$_3$ inversion lines).
The absorption dip at this velocity is more prominent
in the H$_2$O($1_{1,1}-0_{0,0}$) profile as expected from the high critical
density of this transition. 
We disregarded this velocity range in our analysis because of the unknown 
line intensity of the dense core emission.
In contrast, the absorption at $v> 7 $\kms\ corresponds to the gas that is most 
likely to be associated with the W75N cloud in the Cygnus X complex.
Its velocity coverage is very similar in the \CHp\ and \Cp\ spectra. 
We fitted three components in this velocity range, guided by the absorption features
of the HCO$^+$ \Jone\ and H$_2$O profiles
 at $v_{LSR} = 7.5, 11$, and 13.7 \kms shown in Fig. 1, 
the resulting profiles being in red.


The \CHp\ column density of the emission line was determined by assuming 
that it is optically thin. This is
consistent with the \thCHp(1-0) line not being
detected in emission towards galactic massive star-forming regions
\cite[Falgarone et al. in prep., ][]{falgarone05}.  The total \CHp\ column density was
therefore assumed to scale with the line integrated area (in K\kms) with a
coefficient that depends on the excitation temperature. For
40~K$<T_{ex}<100$~K, the range of $T_{ex}$ that minimizes $N(\CHp)$, 
a lower limit $N_{em}(\CHp)=7 \times 10^{11} \pscm\ \int T(v) dv$ was obtained, 
using the spontaneous decay rate $A_{10}=5.9 \times 10^{-3}$ s$^{-1}$ inferred from
the dipole moment, $\mu=1.62$~D \cite{kobayashi93}.  In absorption, we inferred
the column density from the integral of the optical depth (in \kms) to be
$N_{abs}(\CHp)= 3.3 \times 10^{12} \pscm\ \int \tau(v) dv$. It is almost independent
of $T_{ex}$ as long as $T_{ex}<<h\nu/k=39$K, an approximation that is valid for gas
of density much lower than the critical density  of the transition ($>10^6$\cc) and 
not closely associated with intense FIR radiation.  
The resulting \CHp\ column densities  are given in
Table 2. Those of the saturated absorption components are lower limits. 
Although the Gaussian decomposition 
is by no means unique, it provides an estimate of both the detected column densities
and the \CHp\ abundance relative to the 
total hydrogen [\CHp]/[H] $> 1.2 \times 10^{-8}$ in the diffuse gas.
We note that the [C{\sc II}] line opacity of the foreground gas, averaged over 10\kms,  
is $\tau_{\rm [C{\sc II}]}>1.8$. This opacity limit corresponds to
a \Cp\ column density larger than $2 \times 10^{18}$ cm$^{-2}$ 
for gas of density lower than the 
critical density of the transition
$n_{cr}(\rm {C^+})$, i.e. $\sim$ a few 10$^3$\cc, whatever the line 
excitation temperature \citep{crawford85}. 
The required \Cp\ column density is much larger 
if the transition is thermalized, i.e. $n>>n_{cr}(\rm {C^+})$. 
From the estimated
gas column of $1.4 \times 10^{22}$ cm$^{-2}$, the expected \Cp\ column
is $\sim 2 \times 10^{18}$ cm$^{-2}$, assuming a gas-phase carbon abundance
of $1.4 \times 10^{-4}$. This value is in close agreement with that
derived for the low density limit, strengthening the association
of the foreground absorption with  gas of densities lower than 
$\sim$ a few 10$^3$\cc.

\begin{table}
\caption{CH$^+$ Column densities}
\label{tab:col}
\begin{tabular}{lcc}
\hline
Component & $T_{ex}$ & $N(\CHp)$ \\
          & K     & cm$^{-2}$\\
\hline
Broad line emission   &   $40 {\rm K}< T_{ex} < 100{\rm K} $ & $2.2 \times 10^{13}$ \\  
Foreground absorption& 3.0 &  $>1.7 \times 10^{14}$ \\
\hline 
\end{tabular}
\end{table}
%
%
\section{Comparison with model predictions}

  The only formation route of the molecular ion CH$^+$ is understood to be initiated by the
  reaction C$^+$~+~H$_2~\rightarrow$~CH$^+$~+~H, which is highly
  endoenergic ($\Delta E/k=$4640~K). The formation of \CHp\ in the
  cold interstellar medium therefore requires suprathermal energy. 
   Several scenarios have been investigated : C-shocks \citep{P86}, highly illuminated 
  and dense photon-dominated regions
  (PDR) where \Cp\ reacts with vibrationally excited \HH\
  \citep{SD95,agundez10}, turbulent interfaces between the warm and cold neutral medium
  \citep{L07}, and regions of intermittent turbulent dissipation
  \citep[TDR models,][]{godard09}. In the first and last models, ion-neutral friction
plays a major role.

For the DR~21 core, it is unclear which processes are at work
in the \CHp\ formation.  
Broad ($\Delta v=20$~\kms) emission lines of SiO \cite{motte07} 
and vibrationally excited \HH\
\cite{cruz07}, similar to the broad emission in the HIFI profiles, are observed 
and suggest the influence of a shock.
We now present a  preliminary attempt to account
for the observed column densities of CH$^+$ in this highly irradiated environment. 

\subsection{\CHp\ emission line: C-shock models}
We used a steady-state model for C-shocks with time-dependent chemistry and ionization 
\citep{FP03}. The chemical network was
supplemented with the relevant photo-reactions 
(Pineau des For\^ets, priv. comm.).  The \HH\ and CO  
self-shielding were assumed to be those of a PDR model at $n_H=10^4$~cm$^{-3}$.
  We adopted a pre-shock density $n_H=10^4$~cm$^{-3}$, 
a shock velocity
 $v_s$=20 \kms, a magnetic field $B=200 \mu$G \citep[from the measure
  of magnetic field in][]{roberts}, and a standard cosmic-ray 
ionisation rate of $\zeta=5~\times 10^{-17}$s$^{-1}$.  We stopped the
  computation of the shocks at a neutral flow time of 10$^4$~yr, which
  corresponds to the estimated dynamical age of the outflow (lobes of
  0.5~pc for a jet velocity up to 100~\kms). The
  CH$^+$ column-density was found to have already reached a plateau at this stage in
  the shock, so the \CHp\ results are insensitive to that choice of age.

\begin{figure}
  \centering
\includegraphics[width=0.25\textwidth]{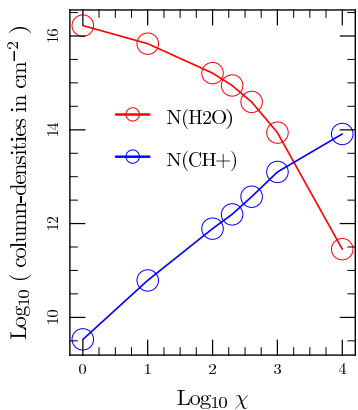}
\caption{\CHp\ (blue) and \wat\ (red)  column densities 
integrated through the C-shock model (see text) for a broad range of UV-field irradiations $\chi$.
The shielding of the shocked material is $A_v=1$.
We note the opposite behaviours of \CHp\ and \wat\ with $\chi$.}
\label{mod_chp}
\end{figure}

Although the incident radiation field is estimated to be 
$\chi=10^5$ in ISRF units \cite{ossenkopf1} for the material closest to the HII region,   
we keep this value as a free parameter.
  Figure \ref{mod_chp} displays the total column densities of 
CH$^+$ and H$_2$O as a function of the adopted external radiation
field. 
Its effect turns out to be quite interesting: a
strong radiation field ($\chi>10^3$) is needed to account for
the large observed CH$^+$ column density. 
It is the enhanced photo-dissociation of \CHtwp\ and \CHthp, both products of the rapid 
hydrogenation of \CHp\ in the shock, that increases the abundance of \CHp\ as $\chi$ increases.
The shock model also predicts that both \CHp\ and \wat\ are
rather well distributed across the various
velocity slices of the shock. This supports the assumption 
that the \CHp\ and \wat\ blue line wings are similar, underlying the 
\CHp\ line frequency determination.

\subsection{\CHp\ absorption lines: turbulent dissipation models}

The large \CHp\ column density observed in the foreground component
corresponds to a {\it los} average of the \CHp\ abundance, 
[\CHp]/[H]$_{ave}>1.2 \times 10^{-8}$, for $N({\rm H})=1.4 \times 10^{22}$ \cq\ (see Sect. 2). 
It can be reproduced in the framework of TDR models. At any time, 
a very large number of tiny regions ($l \sim 100$ AU), altogether 
filling a small fraction of the entire {\it los}, are developing a transient 
warm chemistry triggered by dissipation bursts where local 
\CHp\ abundances reach $10^{-6}$ \citep{godard09}. The filling factor of the ensemble
of these tiny structures
is set by the energy transfer rate in the turbulent cascade, $\epsilon=\rho v_l^3/l$, 
identified with the turbulent dissipation rate.   
The resulting average abundance is found to scale as
[\CHp]/[H]$_{ave}$=6.4$\times 10^{-9} (\epsilon/\epsilon_0) (n_{\rm H}/50\cc)^{-2.6}$
for an ambient radiation field $\chi=3$ in ISRF units,
where $n_{\rm H}$ is 
the density of the gas in which the bursts occur
and $\epsilon$ is a non-local quantity of galactic average $\epsilon_0=2 \times 10^{-25}$ \eccs\ 
\citep[see][]{godard09}. 
Fluctuations by two orders of magnitude are observed in the ISM 
about that galactic average, the highest values 
being reached in active star-forming regions, such as Cygnus X. We thus expect 
$\epsilon$ to be up to 100 times higher than average in Cygnus X, 
so that the observed \CHp\ abundance   
can be produced by intermittent dissipation of turbulence occurring in gas 
of density up to $\sim 250$ \cc.
The total number of tiny structures along that 1.7 kpc {\it los} is found to scale as
$\sim 3.5 \times 10^{3} (\epsilon/\epsilon_0)(n_{\rm H}/50\cc)^{-2}$.

   
%
%
%
\section{Conclusion}
We have detected the line profile of CH$^+$(1-0) towards the DR~21 massive star-forming region  with
the HIFI instrument, obtaining several major results. 
The line is a combination of broad emission and almost saturated broad 
absorption, which have comparable 
integrated areas. The rest-frame frequency 
of the \CHp\Jone\ line is inferred to be $\nu=835137\pm 3$~MHz. 
For the gas seen in emission, the \CHp\ column density,  of about 10$^{13}$ cm$^{-2}$, 
 compares well to predictions of C-shock models 
propagating in dense, highly illuminated gas. 
Additional chemical modelling (and radiative transfer calculations) 
are needed to confirm the C-shock framework and exploit the other line signatures. 
The large CH$^+$ column density in the foreground gas, $N(\CHp) \sim 10^{14}$ cm$^{-2}$,
can be explained by the turbulent dissipation models 
in diffuse gas and confirms the large opacity of that line 
in the diffuse molecular gas, 
%
%

%
%
\begin{acknowledgements}
%
%
HIFI has been designed and built by a consortium of institutes and university
departments from across Europe, Canada and the United States (NASA) under the
leadership of SRON, Netherlands Institute for Space Research, Groningen, The
Netherlands, and with major contributions from Germany, France and the
US. Consortium members are : Canada: CSA, U. Waterloo; France : CESR, LAB,
LERMA, IRAM; Germany : KOSMA, MPIfR, MPS; Ireland : NUI Maynooth; Italy : ASI,
IFSI-INAF, Osservatorio Astrofisico di Arcetri-INAF; Netherlands : SRON, TUD;
Poland : CAMK, CBK; Spain : Observatorio Astron\`omico Nacional (IGN), Centro
de Astrobiologia; Sweden : Chalmers University of Technology - MC2, RSS \&
GARD, Onsala Space Observatory, Swedish National Space Board, Stockholm
University - Stockholm Observatory; Switzerland : ETH Zurich, FHNW; USA :
CalTech, JPL, NHSC.  
MG and EF acknowledge the support from the Centre National de Recherche Spatiale
(CNES). Part of this work was supported by the German \emph{Deuts\-cheFor\-schungs\-ge\-mein\-schaft, DFG} project \# Os177/1--1. We thank G. Pineau des For\^ets for providing us with his version of 
the code for C-shock models. 

\end{acknowledgements}

%
%

\end{document}